\documentclass[conference]{IEEEtran}
\IEEEoverridecommandlockouts
\usepackage{cite}
\usepackage{amsmath,amssymb,amsfonts}
\usepackage{graphicx}
\usepackage{subfigure}
\usepackage{textcomp}
\usepackage{xcolor}
\usepackage{epstopdf}
\usepackage{epsfig}
\usepackage{bm}
\usepackage[linesnumbered,ruled,vlined]{algorithm2e}
\usepackage{algpseudocode}

\usepackage{makecell}
\usepackage{array}
\usepackage{multirow}   
	
\def\BibTeX{{\rm B\kern-.05em{\sc i\kern-.025em b}\kern-.08em
		T\kern-.1667em\lower.7ex\hbox{E}\kern-.125emX}}
	
\begin{document}

\title{Federated Transfer Learning Based Cooperative Wideband Spectrum Sensing with Model Pruning
}

\author{\IEEEauthorblockN{Jibin Jia$^*$, Peihao Dong$^{* \dagger}$, Fuhui Zhou$^*$,  Qihui Wu$^*$}
\IEEEauthorblockA{$^*$ College of Electronic and Information Engineering, Nanjing University of Aeronautics and Astronautics, Nanjing 211106, China\\
	$^\dagger$ National Mobile Communications Research Laboratory, Southeast University, Nanjing 211111, China\\
	Email: \{jiajibin, phdong\}@nuaa.edu.cn, zhoufuhui@ieee.org, wuqihui2014@sina.com}
}

\maketitle

\begin{abstract}
For ultra-wideband and high-rate wireless communication systems, wideband spectrum sensing (WSS) is critical, since it empowers secondary users (SUs) to capture the spectrum holes for opportunistic transmission. However, WSS encounters challenges such as excessive costs of hardware and computation due to the high sampling rate, as well as robustness issues arising from scenario mismatch. In this paper, a WSS neural network (WSSNet) is proposed by exploiting multicoset preprocessing to enable the sub-Nyquist sampling, with the two-dimensional convolution design specifically tailored to work with the preprocessed samples. A federated transfer learning (FTL) based framework mobilizing multiple SUs is further developed to achieve a robust model adaptable to various scenarios, which is paved by the selective weight pruning for the fast model adaptation and inference. Simulation results demonstrate that the proposed FTL-WSSNet achieves the fairly good performance in different target scenarios even without local adaptation samples.
\end{abstract}

\begin{IEEEkeywords}
Cognitive radio, wideband spectrum sensing, sub-Nyquist, weight pruning, federated transfer learning
\end{IEEEkeywords}

\section{Introduction}
With the skyrocketing proliferation of wireless applications, spectrum scarcity has become increasingly prominent, severely limiting the transmission capability of future communication systems \cite{P. Dong}. Cognitive radio (CR) aims to enable SUs to share and utilize the licensed but idle spectrum bands in a dynamic access manner without causing the substantial interference to primary users (PUs), thereby enhancing spectrum utilization and alleviating spectrum scarcity. Wideband spectrum sensing is indispensable for achieving dynamic spectrum access over a wide frequency range in CR systems \cite{b2}\cite{b}.

Deep learning (DL), known as its powerful nonlinear fitting capabilities, autonomously learns the underlying features and has emerged as an effective approach for addressing dynamic decision-making challenges in complex spectrum environments. In \cite{b3}, the discrete signal data obtained by multicoset sampling was fed into a deep neural network (DNN) for WSS. A parallel convolutional neural network (CNN) based WSS architecture was designed in \cite{b4}, which individually processes the sampled data divided into two parts to reduce latency. A Time-Frequency-Fused adjustable deep convolutional neural network (TFF-aDCNN) was pre-trained as a base model for WSS, and then rapidly fine-tuned by using transfer learning (TL) to adapt to the actual environment in \cite{b5}. Nevertheless, these centralized DL approaches necessitate substantial training data, which may not be feasible to obtain in practice due to data privacy concerns.

Recently, federated learning (FL) has attracted considerable attention as a novel paradigm mobilizing multiple devices to collaboratively train a DL model without directly exchanging raw private data, thus effectively ensuring privacy protection \cite{b6} \cite{b7}. Most research works have already explored the application of FL in narrowband spectrum sensing (NSS). In \cite{b8}, FL was conducted to collaboratively train a DNN model over a multi-hop wireless network for spectrum sensing. In \cite{b9}, FL nodes were grouped based on the mean Signal-to-Noise Ratio (SNR) of received data and a shared CNN model was developed for each group to improve the detection performance. A FL framework is first applied to cooperative spectrum sensing, where the server serves as the fusion center to combine the sensing results from each node in \cite{b10}. 

Different from the above centralized DL approaches\cite{b3,b4,b5} and distributed FL-based NSS approaches \cite{b8,b9,b10}, in this paper, a FL based cooperative WSS framework is proposed. Despite inheriting the advantages of FL, the FL-based WSS approaches are also faced with similar challenges in deploying complex models on mobile edge devices  \cite{b11}. Additionally, due to the activity dynamics of PUs working at different frequency bands in wireless environment, they are also plagued by generalization issues. To overcome these problems, we develop a versatile federated transfer learning (FTL) based WSS framework with model pruning, aiming at improving the robustness to different scenarios in a lightweight manner without exposing raw spectrum data. The main innovations and contributions are outlined as follows:

\begin{itemize}[\IEEEsetlabelwidth{Z}]
\item[1)] The low-cost Sub-Nyquist sampling is facilitated by applying multicoset preprocessing, which offers the estimated spectrum  of the wideband signal. Subsequently, the preliminary feature is fed into a designed WSSNet to detect spectrum holes with high accuracy. To reduce the complexity of WSSNet for ease of deployment and adaptation, a selective weight pruning strategy is proposed for the offline-trained WSSNet to remove weights that have negligible impact on performance from the specific layer.

\item[2)] A FTL-based model adaptation mechanism is developed to realize the knowledge interaction among models deployed at different SUs and thus to guarantee the decent detection accuracy for these SUs in more unseen scenarios even without adaptation samples.

\end{itemize}

\section{System Model} \label{System Model}

In this section, we consider a CR system including $K$ PUs and one SU monitoring over a wide frequency range $[0,B]$ for dynamic spectrum access. During the spectrum sensing period, the received signal by the SU can be given by
\begin{align}
\label{eqn_xt}
x(t) = \sum_{k=1}^{K} s_{k}(t) + n(t),
\end{align}
where $s_{k}(t)$ denotes the signal transmitted by the $k$th PU and $n(t)\sim\mathcal{CN}(0,\sigma^2)$ denotes the circular symmetric complex additive white Gaussian noise (AWGN) with variance $\sigma^2$. Performing Fourier transform on $x(t)$ yields
\begin{align}
\label{eqn_Xf}
X(f) = \int_{-\infty}^{\infty} x(t) e^{-j2\pi f t} dt  = S(f) + N(f),
\end{align}
where $S(f)$ and $N(f)$ denote the spectrums of the compound PU signal and AWGN, respectively.

\begin{figure}[t]
\centering
\includegraphics[width=3.2in]{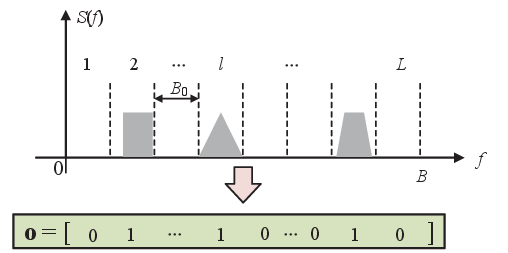}
\caption{Occupation model of the wideband spectrum.}\label{occupation model}
\end{figure}

In the CR, the wideband spectrum of interest is uniformly partitioned into $L$ consecutive and non-overlapped sub-bands, each with the bandwidth $B_{0}=B/L$ and indexed from 1 to $L$, as shown in Fig.~\ref{occupation model}. Given that the signal of each PU randomly occupies one of the $L$ sub-bands, the wideband spectrum occupancy detection can be modeled as multiple binary classification problems. Then the spectrum occupancy situation can be indicated by an $L$-entry binary vector, $\mathbf{o}$, which contains $K$ non-zero elements. 

Given the dynamic nature of licensed spectrum resource allocated to PUs, the SU must conduct periodic spectrum sensing to seize more transmission opportunities. However, traditional WSS approaches are constructed based on Nyquist sampling rate, causing the unaffordable hardware cost and power consumption. Then we aim to design an efficient neural network architecture to realize WSS with sub-Nyquist sampling rate in the following sections.

\section{WSSNet with Sub-Nyquist Sampling}

In this section, our main focus is on the design of the WSSNet architecture. Before delving into that, we exploit a multicoset preprocessing procedure with sub-Nyquist rate to obtain an intermediate feature facilitating the training of WSSNet.

\subsection{Multicoset Preprocessing}
Define the Nyquist frequency as $f_{\textrm{N}} = 1/T$, where $T$ is the Nyquist sampling period. Multicoset sampling is a periodic non-uniform sub-Nyquist sampling technique, which employs $P$ parallel cosets, each corresponding to uniformly sample the signal $x(t)$ at a sub-Nyquist rate $f_{\textrm{e}}=1/LT$ with a unique time delay $c_pT$. With $c_p \in \mathbb{Z}$ satisfying 
\begin{align}
\label{eqn_c_p}
0 \leq c_1<\cdots< c_p <\cdots<c_P\leq L-1, 
\end{align}
the set $\mathcal{C}=\{c_p\}_{p=1}^P$ guarantees to obtain $P$ samples within sampling interval $LT$, yielding the sampling rate $f_{\textrm{s}}=\frac{P}{LT}=\frac{P}{L}f_{\textrm{N}}$. Based on the prior information on the number of sub-bands $L$, the SU adopts a $P$-coset sampler to enable sub-Nyquist sampling, provided that $P<L$ is set. In the $p$th coset, the sampled sequence can be expressed as
\begin{align}
\label{eqn_y_p}
y_p[n] = x(nLT+c_p T). \quad n\in \mathbb{Z},
\end{align}
The corresponding discrete-time Fourier transform (DTFT) is denoted as $Y_p(e^{j2\pi fLT})$. Define $y_p(f)=e^{-j2\pi f c_p T} Y_p(e^{j2\pi fLT})$ and $x_{l}(f)=X(f-\frac{l-1}{LT})$. Then stacking them for all $ p=1,\dots,P$ and $l=1,\dots,L $, respectively, can establish the relationship $\mathbf{y}(f)=\mathbf{A} \mathbf{x}(f)$ for $f \in [0,B/L)$, where $\mathbf{A}\in\mathbb{C}^{P\times L}$ is a measurement matrix with the entry $A_{p,l}=\frac{e^{-j2\pi (l-1) c_p/L}}{LT} $ \cite{b3}.
Considering practical applications, discrete Fourier transform (DFT) is often used to replace DTFT \cite{b12}. Given the $N$-point sampled sequence acquired by each coset, the $P$-coset sampler can generate a matrix form, that is
\begin{align}
\label{eqn_y_matrix}
\mathbf{Y}=\mathbf{A} \mathbf{X},
\end{align}
where $\mathbf{Y}\in\mathbb{C}^{P\times N}$ and $\mathbf{X}\in\mathbb{C}^{L\times N}$. Then $\mathbf{X}$ can be roughly recovered from $\mathbf{Y}$, that is 
\begin{align}
\label{eqn_x_matrix_hat}
\hat{\mathbf{X}}=\mathbf{A}^{\dagger}\mathbf{Y}=\mathbf{A}^{\dagger}\mathbf{A} \mathbf{X},
\end{align}
where $(\cdot)^{\dagger}$ denotes the pseudo-inverse. The normalized matrix $\bar{\mathbf{X}}=\frac{\hat{\mathbf{X}}}{|\hat{\mathbf{X}}|}$ serves as a preliminary feature input to the subsequent WSSNet for predicting spectrum occupancy.

\subsection{WSSNet Architecture and Training}

Considering that the WSS task can be transformed into a multi-label binary classification task, the function of WSSNet mainly includes feature extraction and class recognition. Thus, the combination of two-dimensional (2D) convolutional layers and fully-connected (FC) layers is adopted for WSSNet. The first two 2D convolutional layers have $32$ and $16$ kernels of size $3\times 3$, respectively. The subsequent FC layer has $128$ neurons and the output layer has $L$ neurons. An $L\times N\times 2$ tensor, consisted of the real and imaginary parts of the feature matrix $\bar{\mathbf{X}}$, is fed into WSSNet to output the soft prediction of the occupancy vector, $\tilde{\mathbf{o}} \in \mathbb{R}^{L\times1}$. Each hidden layer utilizes the rectified linear unit (ReLU) activation function, and the output layer employs the Sigmoid activation function to obtain the probability scores. Furthermore, dropout is applied to each hidden layer to avoid overfitting.

WSSNet is trained with a labeled dataset expressed as
\begin{align}
\label{eqn_dataset}
\mathcal{D}_{\textrm{S}}=\{(\bar{\mathbf{X}}^{(1)}, \mathbf{o}^{(1)}),\dots,(\bar{\mathbf{X}}^{(n)}, \mathbf{o}^{(n)}),\dots,(\bar{\mathbf{X}}^{(N_{\textrm{tr}})}, \mathbf{o}^{(N_{\textrm{tr}})})\},
\end{align}
where $N_{\textrm{tr}}$ denotes the number of training samples and the superscript $(n)$ indicates the $n$th sample. Considering that the learning task is formulated as multiple binary classification problems, the binary cross-entropy is used as the loss function to measure the discrepancy between the prediction and label vectors in an element-wise manner, that is,
\begin{align}
\label{eqn_loss}
\mathcal{L} = -\frac{1}{N_{\textrm{tr}}}\sum_{n=1}^{N_{\textrm{tr}}}\sum_{l=1}^{L} o_{l}^{(n)}\log \tilde{o}_{l}^{(n)} + (1-o_{l}^{(n)})\log (1-\tilde{o}_{l}^{(n)}),
\end{align}
where $\tilde{o}_{l}^{(n)} $ denotes the $l$th element of $\tilde{o} $ for the $n$th sample. The goal of model training is to obtain optimal parameters by minimizing the loss function $\mathcal{L} $ over $\mathcal{D}_{\textrm{S}}$. Using backpropagation algorithm, the model parameters are progressively updated by
\begin{align}
\label{eqn_update}
\boldsymbol{\Theta}\gets \boldsymbol{\Theta}-\alpha\nabla \mathcal{L}_{\mathcal{D}_{\mathrm{S}}}(\boldsymbol{\Theta}),
\end{align}
where  $\alpha$ is the learning rate and $\nabla \mathcal{L}_{\mathcal{D}_{\mathrm{S}}}(\cdot )$ is the loss gradient based on the dateset $\mathcal{D}_{\mathrm{S}}$. This process is repeated until the validation loss no more
decreases, yielding a decent base model for the subsequent online adaptation and inference.

\section{Lightweight and Robust WSSNet Design via FTL}
\begin{figure}[t]
\centering
\includegraphics[width=3.5in]{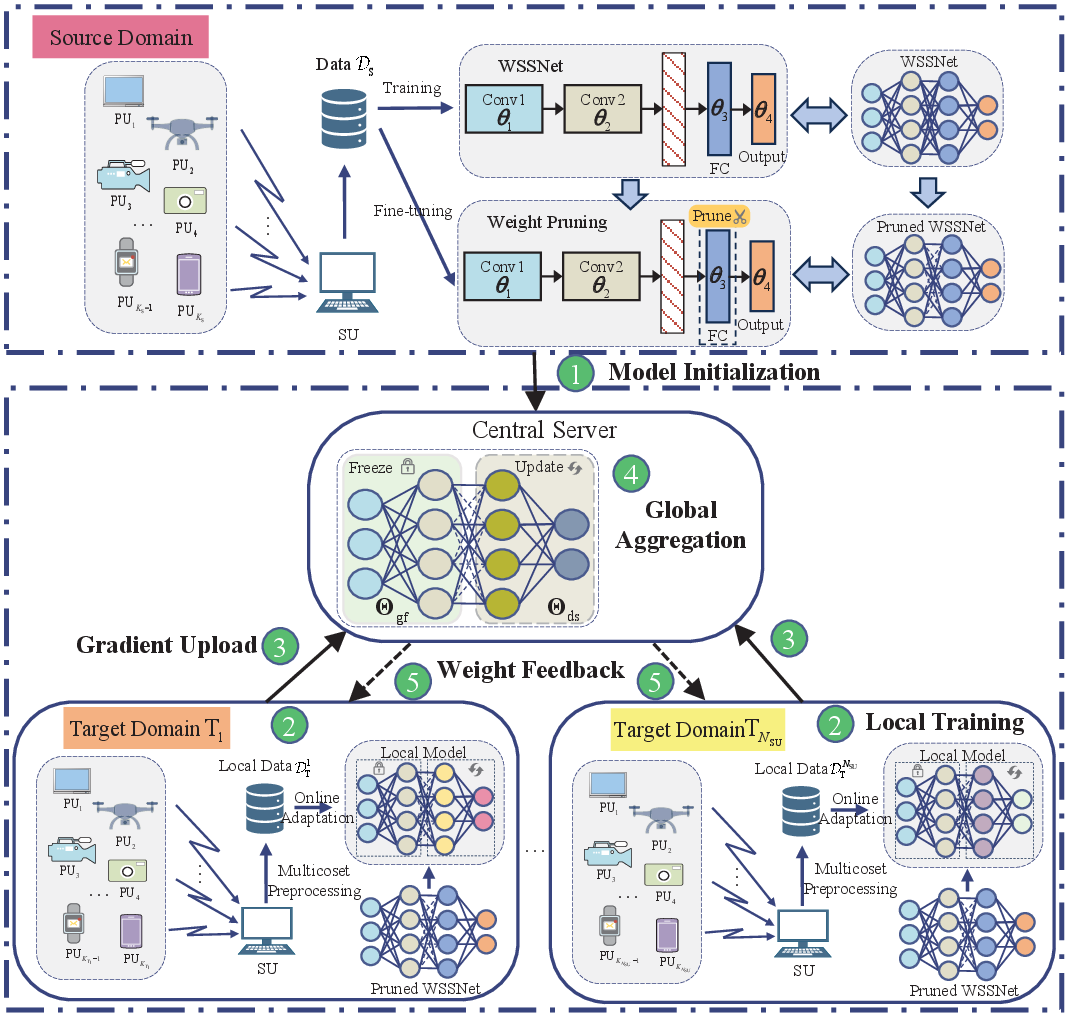}
\caption{Workflow of FTL-WSSNet.}\label{FTL-WSSNet}
\end{figure}

As the spectrum environments complicate, the size of WSSNet may be enlarged sharply. Consequently, the computation, storage, and power resources used for model training and inference increase significantly, restricting the deployment of WSSNet on mobile edge devices with the limited resources and low-cost hardware. Moreover, the activity of PUs on the spectrum bands is dynamic, in terms of the positions and the number of occupied sub-bands, posing challenges for the generalization capability of WSSNet. While WSSNet can relatively easily cope with the variation in the positions of occupied sub-bands, grasping the varying number of occupied sub-bands, or equivalently occupancy ratios, is quite challenging. To tackle the above issues, a FTL-based framework is developed by mobilizing multiple SUs deploying WSSNet, in order to accelerate the online model adaptation and inference significantly as well as reduce the model size. The workflow of the FTL-based WSSNet (FTL-WSSNet) is depicted in Fig.~\ref{FTL-WSSNet} and will be elaborated in the following three subsections.

\subsection{Weight Pruning}
The original WSSNet suffers from the problem of large number of parameters, due to a lot of redundant neural connections inside. Therefore, weight pruning can be employed to remove insignificant weights without compromising the prediction accuracy. The reduction of the quantity of model weights can facilitate the deployment of WSSNet on mobile devices and accelerate the online model adaptation. Denote the weight set of WSSNet as $\boldsymbol{\Theta}=\{\bm{\theta}_{1},\bm{\theta}_{2},\bm{\theta}_{3},\bm{\theta}_{4}\}$. Specially, we opt to apply weight pruning to $\bm{\theta}_{3}$ corresponding to the FC layer containing the most of weights of WSSNet and skip the convolutional layers with quite fewer weights but crucial to the performance, as shown in Fig.~\ref{FTL-WSSNet}. Magnitude-based weight pruning is applied to remove weights with absolute values less than a predefined threshold. All $N$ weights in $\bm{\theta}_{3}$ are sorted in ascending order based on their absolute values to form a vector $\bm{\eta}$. Given the pruning ratio $\kappa$, the pruning threshold $\gamma$ can be determined by
\begin{eqnarray}
\label{eqn_threshlod}
\gamma = [\bm{\eta}]_{\lceil\kappa{N}\rceil},
\end{eqnarray}
where $\lceil\cdot\rceil $ and $[\bm{\eta}]_i$ denote the ceiling function and the $i$th element of $\bm{\eta}$, respectively. Then the weights in $\bm{\theta}_3$ can be processed as
\begin{align}
\label{eqn_prun}
[\bm{\theta}_3]_i =
\begin{cases}
	[\bm{\theta}_3]_i, &  |[\bm{\theta}_3]_i| \geq \gamma \\
	0, & \text{otherwise}
\end{cases},
\end{align}
where $|\cdot|$ denotes the absolute value. After pruning, the complexity of WSSNet is reduced, accompanied by some degree of performance degradation. Therefore, the final step involves fine-tuning the remaining weights over several epochs using the dataset $\mathcal{D}_{\textrm{S}}$ and loss function $\mathcal{L}$ to compensate the performance loss.

\subsection{Model Adaptation}
In TL, knowledge learned from one data distribution (source domain) is transferred to another data distribution (target domain) \cite{b13}. FTL enhances TL by invoking multiple devices using the same model to share the knowledge learned from their respective new scenarios, and thus is suitable for the CR network including several geographically distributed SUs described in Section \ref{System Model}. Specifically, it enables different devices or nodes to train models locally and share model parameters or feature representations to facilitate knowledge transfer across them. Compared to TL, FTL brings benefits in two folds: 1) Improve model's generalization ability by involving multiple SUs owning respective spectrum data with different data distribution or characteristics to rich the virtual data pool. 2) Protect data privacy by exchanging weight gradients representing the spectrum knowledge among SUs instead of sharing raw data. 

Consider a CR network consisted of $N_\mathrm{SU}$ geographically distributed SUs monitoring the same spectrum bands. We propose a FTL-based online model adaptation algorithm to improve WSSNet trained in the source domain so that it is still effective in the unseen target domain, as illustrated in Fig.~\ref{FTL-WSSNet}. The overall process of the algorithm consists of  five steps.

\begin{itemize}
\item[\emph{1)}] \emph{Model Initialization:} The central server uses the pruned WSSNet trained in the source domain as the initial global model and broadcasts the model parameters $\boldsymbol{\Theta}$ to SUs.
\item[\emph{2)}] \emph{Local Training:} SUs utilize their own data in the target domain to perform the local model update based on the initial parameters $\boldsymbol{\Theta} $. Consider that the layers of WSSNet can be divided into the general-feature layers and the domain-specific layers, corresponding to the weight sets $\boldsymbol{\Theta}_{\mathrm{gf}}=\{\bm{\theta}_{1},\bm{\theta}_{2}\}$ and $\boldsymbol{\Theta}_{\mathrm{ds}}=\{\bm{\theta}_{3},\bm{\theta}_{4}\}$, respectively, with $\boldsymbol{\Theta}_{\mathrm{gf}}\cup \boldsymbol{\Theta}_{\mathrm{ds}}=\boldsymbol{\Theta}$. Therefore, the update strategy is to freeze $\boldsymbol{\Theta}_{\mathrm{gf}}$ for transferring the general features of the source domain and to update $\boldsymbol{\Theta}_{\mathrm{ds}}$ to fit the new target domain.
\item[\emph{3)}] \emph{Gradient Upload:} After completing the local model update of each batch, SUs upload the accumulated gradients in the training process to the central server. Only the accumulated gradient $\boldsymbol{G}_{\mathrm{ds}}^{t,i}$ corresponding to $\boldsymbol{\Theta}_{\mathrm{ds}}^{t,i}$ is uploaded to the central server, where $t$ and $i$ denote the $t$th communication round and the $i$th SU, respectively. Compared to uploading all gradients $\boldsymbol{G}^{t,i}$, the operation reduces the communication overhead and improves the FL efficiency.
\item[\emph{4)}] \emph{Global Aggregation:} The central server aggregates the gradients collected from various SUs, and utilizes the aggregated gradients to update the global model by using the stochastic gradient descent.
\item[\emph{5)}] \emph{Weight Feedback:} The server broadcasts the updated parameters to each SU for the next iteration. 
\end{itemize}
The last four steps described above are repeated until all communication rounds have been completed. The details of the process are summarized in Algorithm \ref{alg1}. The collaborative online adaptation process yields a more general model, FTL-WSSNet, better fitting the complex and dynamic spectrum environment due to the capability of accommodating diverse scenarios.  When a SU encounters with a new scenario that has been seen by other SUs, FTL-WSSNet is able to achieve the satisfactory performance even without adaptation samples, so that the zero-shot adaptation can be realized.
\begin{algorithm}[t]  
\caption{FTL-based Online Model Adaptation}
\label{alg1}
\LinesNumbered  
\KwIn{local dataset in the taget domain $\mathcal{D}_{\mathrm{T}}^i $, number of SUs $N_{\mathrm{SU}}$, number of communication rounds $M$, number of local epochs $E$, number of local batch size $b$, learning rate $\alpha $, weight of pruned WSSNet $\boldsymbol{\Theta}=\boldsymbol{\Theta}_{\mathrm{gf}}\cup \boldsymbol{\Theta}_{\mathrm{ds}}$}  
\KwOut{The final global model $\boldsymbol{\Theta}^M$}
\textbf{Server executes:}

Initialize: $\boldsymbol{\Theta}^0 \gets \boldsymbol{\Theta}$  \\
$\boldsymbol{\Theta}^0 = \boldsymbol{\Theta}_{\mathrm{gf}}^0 \cup \boldsymbol{\Theta}_{\mathrm{ds}}^0$ 

\For{ $t=0,1,\dots,M-1$ }  
{$N_{\mathrm{ad}} \gets \sum_{i=1}^{N_{\mathrm{SU}}}|\mathcal{D}_{\mathrm{T}}^i|$
	
	\For{$i=1,2,\dots,N_{\mathrm{SU}}$ \textup{\textbf{in parallel}}}{
		send the global model $\boldsymbol{\Theta}^t$ to $i$th SU
		
		$\boldsymbol{G}_{\mathrm{ds}}^{t,i} \gets$ \textbf{LocalTraining($i,\boldsymbol{\Theta}^t$)}}
	$\boldsymbol{\Theta}_{\mathrm{ds}}^{t+1} \gets \boldsymbol{\Theta}_{\mathrm{ds}}^{t}-\alpha\sum_{i=1}^{N_{\mathrm{SU}}}\frac{|\mathcal{D}_{\mathrm{T}}^i|}{N_{\mathrm{ad}}}\boldsymbol{G}_{\mathrm{ds}}^{t,i}$
	
	$\boldsymbol{\Theta}^{t+1} \gets  \boldsymbol{\Theta}_{\mathrm{gf}}^0\cup \boldsymbol{\Theta}_{\mathrm{ds}}^{t+1}$
	
}  
\textbf{return} $\boldsymbol{\Theta}^M$

\textbf{SU executes:}

\textbf{LocalTraining}($i,\boldsymbol{\Theta}^t$):

$\boldsymbol{\Theta}^{t,i} \gets \boldsymbol{\Theta}^{t},  \boldsymbol{G}_{\mathrm{ds}}^{t,i} \gets 0$ \\
$\boldsymbol{\Theta}^{t,i} = \boldsymbol{\Theta}_{\mathrm{gf}}^0 \cup \boldsymbol{\Theta}_{\mathrm{ds}}^{t,i} $

\For{$e=0,1,\dots,E-1$}
{
	\For{\textup{each batch} $b \in \mathcal{D}_{\mathrm{T}}^i$}
	{compute $\mathcal{L}_b(\boldsymbol{\Theta}^{t,i})$ according to (\ref{eqn_loss})
		
		$\boldsymbol{\Theta}_{\mathrm{ds}}^{t,i}\gets \boldsymbol{\Theta}_{\mathrm{ds}}^{t,i}-\alpha\nabla \mathcal{L}_{b}(\boldsymbol{\Theta}^{t,i})$
		
		$\boldsymbol{G}_{\mathrm{ds}}^{t,i}\gets \boldsymbol{G}_{\mathrm{ds}}^{t,i}+\nabla \mathcal{L}_{b}(\boldsymbol{\Theta}^{t,i})$
		
		$\boldsymbol{\Theta}^{t,i} \gets \boldsymbol{\Theta}_{\mathrm{gf}}^0 \cup \boldsymbol{\Theta}_{\mathrm{ds}}^{t,i}$
	}
}
\textbf{return} $\boldsymbol{G}_{\mathrm{ds}}^{t,i}$ to the server

\end{algorithm}
\subsection{Model Inference}
After weight pruning and model adaptation, a lightweight and robust FTL-WSSNet deployed on SUs is used for spectrum occupancy prediction. The inference phase is performed independently by each SU to verify that the model is effective for different scenarios. Each SU collects local data with  multicoset preprocessing and then feeds them into the network to obtain the soft prediction of the occupancy vector, $\tilde{\mathbf{o}}$. By comparing each element of $\tilde{\mathbf{o}}$ with a given threshold $\lambda \in (0,1)$, the final prediction vector $\hat{\mathbf{o}}$ with binary elements is obtained to indicate the spectrum occupancy situation, that is 
\begin{align}
\label{eqn_inference}
\hat{o}_l =
\begin{cases}
	1, &  \tilde{o}_l \geq \lambda \\
	0, & \tilde{o}_l < \lambda
\end{cases},
\end{align}
where $\tilde{o}_l$ and $\hat{o}_l$ correspond to the $l$th element of $\tilde{\mathbf{o}}$ and $\hat{\mathbf{o}}$, respectively. Based on this, each SU can periodically perform spectrum sensing to be aware of the surrounding spectrum environment, enabling dynamic spectrum access.

\section{Numerical Simulations}

\subsection{Simulation Setup}
Consider a WSS system over a wide frequency range $[0,320]$ MHz containing $L=40$ non-overlapped sub-bands, each with the bandwith $B_0=8$ MHz. The received signal can be generated by 
\begin{eqnarray}
\label{eqn_st}
x(t) = \sum_{k=1}^{K} \sqrt{E_kB_0} \operatorname{sinc}(B_0(t-t_k))e^{j2\pi f_kt} + n(t)
\end{eqnarray}
where $\operatorname{sinc}(x)=\sin(\pi x)/(\pi x)$, $E_k$, $t_k$ and $f_k$ are the energy, time offset and carry frequency of $k$th PU's signal, respectively. Set $E_k=1$, $t_k\in{(0,T_d)}$ and $f_k\in{[B_0/2,B-B_0/2]}$, where $T_d=8$ $\mu s$ is the duration time of signal. In the multicoset preprocessing stage, the number of cosets is $P=8$, each with $N=64$ sampled data. In the offline training stage, the numbers of training, validation, and testing samples are $N_{\text{tr}}=12000$, $N_{\text{va}}=4000$, and $N_{\text{te}}=4000$, respectively. In the weight pruning stage, the pruning ratio is $\kappa=0.9$. In the model adaptation stage, the number of SUs is $N_{\text{SU}}=4$ with adaptation samples $N_{\text{ad}}^i=100$, for $i=1,2,3,4$. In the model inference stage, the decision threshold is $\lambda=0.5$. To evaluate the FTL-WSSNet performance, the prediction accuracy, $P_{\mathrm{acc}}=\frac{1}{N_{\mathrm{te}}}\sum_{i=1}^{N_{\mathrm{te}}}\frac{N_{\mathrm{co}}^i}{L}$, is used as the performance metric, where $N_{\mathrm{co}}^i$ denotes the number of sub-bands predicted correctly for the $i$th testing sample. Assuming there are four target domains, $\text{T}_1 $, $\text{T}_2 $, $\text{T}_3 $, and $\text{T}_4 $, each with one SU. To simulate the different spectrum environments in each domain, we set the number of PUs in these four domains to be $K_{\text{T}_1}=8$, $K_{\text{T}_2}=12$, $K_{\text{T}_3}=16$, and $K_{\text{T}_4}=24$, respectively. In the source domain, the number of PUs is $K_\text{S} = 20$. 

\subsection{Simulation Results}
\begin{figure}[t!]
\centering
\includegraphics[width=3.2in]{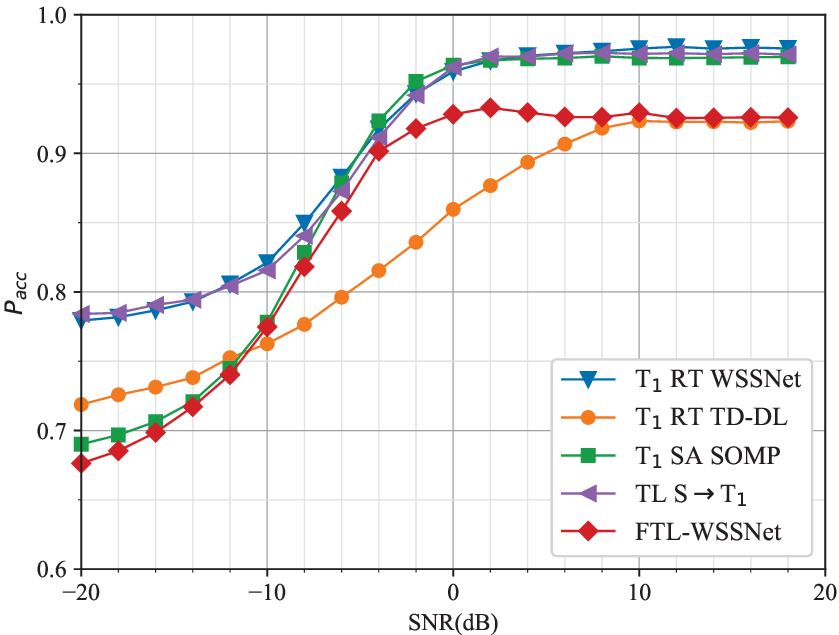}
\caption{Prediction accuracy versus SNR in the target domain ${\text{T}_1}$.}\label{fig4}
\end{figure}
\begin{figure}[t!]
\centering
\includegraphics[width=3.2in]{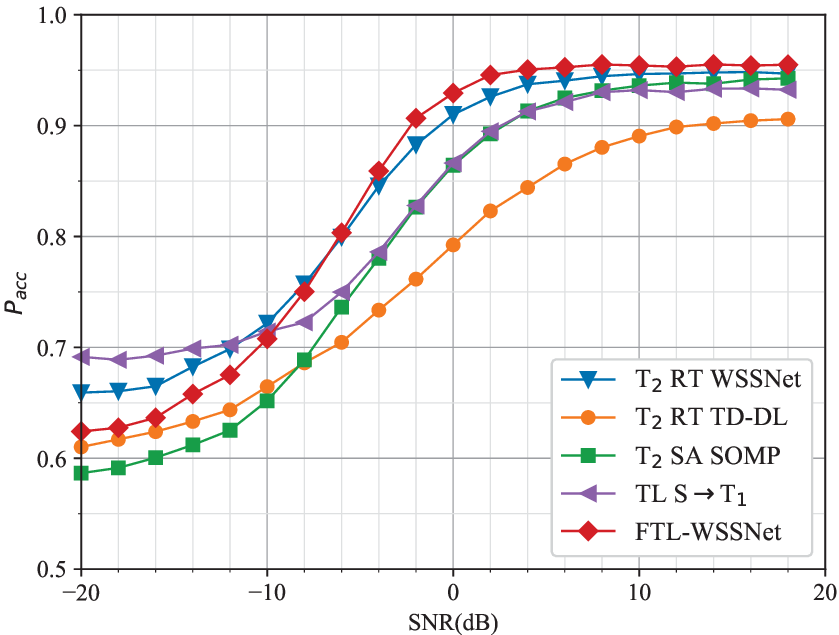}
\caption{Prediction accuracy versus SNR in the target domain ${\text{T}_2}$.}\label{fig5}
\end{figure}

\begin{figure}[t!]
\centering
\includegraphics[width=3.2in]{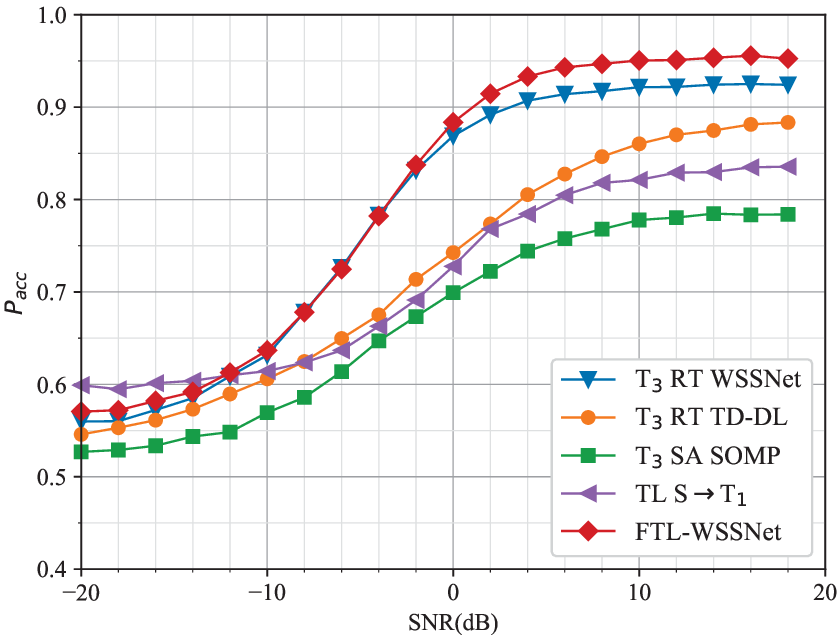}
\caption{Prediction accuracy versus SNR in the target domain ${\text{T}_3}$.}\label{fig6}
\end{figure}

\begin{figure}[t!]
\centering
\includegraphics[width=3.2in]{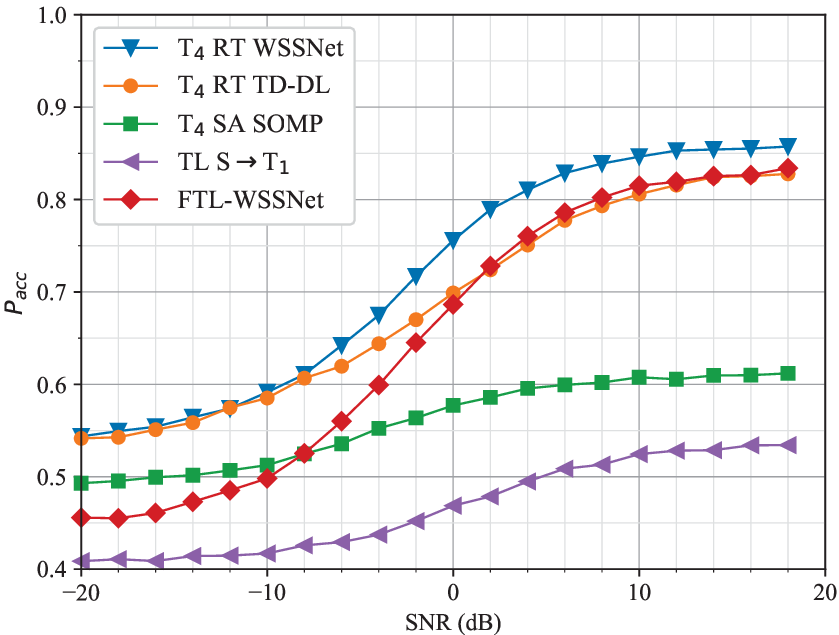}
\caption{Prediction accuracy versus SNR in the target domain ${\text{T}_4}$.}\label{fig7}
\end{figure}

The time domain-based DL (TD-DL) scheme in \cite{b3}, and the simultaneous orthogonal matching pursuit (SOMP) scheme in \cite{b14} act as baselines for performance comparison. Fig.~\ref{fig4}-\ref{fig7} respectively show the prediction accuracy versus SNR in the target domain ${\text{T}_i}$, where $i=1,2,3,4$. Specifically, ``${\text{T}_i}$ RT WSSNet" and ``${\text{T}_i}$ RT TD-DL" denote the regular training mode with sufficient training samples from the matched domain ${\text{T}_i}$. The complexities are $\mathcal{O}\left(\sum_{i=1}^{2}D_{i,1}D_{i,2}N_{i}+\sum_{i=3}^{4}N_{i}\right)$ and $\mathcal{O}\left(\sum_{i=1}^{3}M_{i}\right)$, respectively, where $D_{i,1}$ and $D_{i,2}$ denote the length and width of output feature maps in the $i$th layer, and $N_i$ and $M_i$ denote the number of parameters in the $i$th layer of their respective networks. ``${\text{T}_i}$ SA SOMP" refers to the sparsity-aware SOMP scheme in the target domain ${\text{T}_i}$, which has a complexity of $\mathcal{O}\left(K^3L\right)$. ``TL S $\rightarrow \text{T}_1$" denotes the transfer training with the adaptation samples from the target domain $\text{T}_1 $ by using the pruned WSSNet trained in the source domain as an initial model, and ``FTL-WSSNet" refers to the proposed FTL-based online model adaptation scheme. With weight pruning, the complexities of the two schemes are $\mathcal{O}\left(\sum_{i=1}^{2}D_{i,1}D_{i,2}N_{i}+(1-\kappa)N_{3}+N_4\right)$. Next, we analyze prediction accuracy of the five schemes when the SU in $\text{T}_1 $ encounters with the same scenario $\text{T}_1 $ or a new scenario  ${\text{T}_i}$, where $i=2,3,4$.

Fig.~\ref{fig4} shows the comparison when the SU in $\text{T}_1 $ still encounters with the scenario $\text{T}_1$. ``FTL-WSSNet", although ranking lower, still achieves 0.9527 of the highest accuracy among all schemes. Fig.~\ref{fig5} shows the comparison when the SU in $\text{T}_1 $ encounters with the new scenario $\text{T}_2 $. ``FTL-WSSNet" is optimal and ``TL S $\rightarrow\text{T}_1$" remains considerable, as the number of occupied sub-bands in $\text{T}_1$ and $\text{T}_2$  is close. Fig.~\ref{fig6} shows the comparison when the environment changes from $\text{T}_1 $ to $\text{T}_3 $. ``FTL-WSSNet" clearly outperforms other schemes. Both the performance of ``${\text{T}_3}$ SA SOMP" and ``TL S $\rightarrow\text{T}_1$" significantly decline, with the former attributed to increased occupancy ratio and the latter to increased environmental differences. Fig.~\ref{fig7} shows the comparison when the SU in $\text{T}_1 $ encounters with the new scenario $\text{T}_4 $. Obviously, both ``TL S $\rightarrow\text{T}_1$ and ``${\text{T}_4}$ SA SOMP" schemes are no longer suitable for $\text{T}_4 $. ``FTL-WSSNet" still maintains excellent performance reaching 0.9630 of the highest accuracy. Table~\ref{tab:method-compa} provides a clearer comparison of prediction accuracy at SNR=10 dB in different scenarios, where ``ratio" is the ratio of the accuracy of the current scheme to that of the optimal scheme. 
From the regular training schemes of the four figures, although WSSNet exhibits excellent performance, the mode, which requires collecting a large number of training samples from new environments, is not suitable for online adaptation. The SOMP scheme requires prior knowledge of sparsity and performs poorly when the number of cosets is less than the number of occupied sub-bands. The TL scheme proves to be ineffective, while FTL-WSSNet can achieve the satisfactory performance with zero-shot adaptation, when facing significant environmental changes. FTL-WSSNet achieves a balance between centralized training and TL by obviating the need for aggregating datasets in a central location while maintaining effectiveness across diverse scenarios.

\begin{table}[t!]
\centering
\caption{Comparison of Prediction Accuracy in Different scenarios}
\label{tab:method-compa}

\resizebox{\linewidth}{!}{
	\begin{tabular}{c|cc|cc|cc|cc|cc}
		\Xhline{1pt}
		& \multicolumn{2}{c|}{FTL-WSSNet} & \multicolumn{2}{c|}{$\text{S}\rightarrow \text{T}_1$} & \multicolumn{2}{c|}{RT WSSNet} & \multicolumn{2}{c|}{RT TD-DL} & \multicolumn{2}{c}{SA SOMP} \\
		\Xcline{2-11}{0.5pt}
		& rank & ratio &  rank & ratio  &  rank & ratio & rank & ratio & rank & ratio  \\
		\Xhline{1pt}
		$\text{T}_{\mathrm{1}}$ & 4 & 0.9527 & 2 & 0.9961 & 1 &	1 & 5 & 0.9465 & 3 & 0.9931 \\
		$\text{T}_{\mathrm{2}}$ & 1 & 1 & 4 & 0.9767 & 2 & 0.9920 & 5 & 0.9333 & 3 & 0.9810 \\
		$\text{T}_{\mathrm{3}}$ & 1 & 1 & 4 & 0.8640 & 2 & 0.9695 & 3 & 0.9051 & 5 & 0.8183 \\
		$\text{T}_{\mathrm{4}}$ & 2 & 0.9630 & 5 & 0.6197 & 1 & 1 & 3 & 0.9522 & 5 & 0.7180 \\

		\Xhline{1pt}
	\end{tabular}
}
\end{table}

\section{Conclusion}
In this article, WSSNet is designed to facilitate learning feature from multicoset preprocessed data, enabling sub-Nyquist sampling. Subsequently, weight pruning is applied to WSSNet for deployment and adaptation. The FTL-based online model adaptation mechanism mobilizing multiple SUs is further developed to address the scenario mismatch without exposing raw spectrum data. Simulation results verify that FTL-WSSNet can achieve fairly good performance in different target scenarios even without local adaptation samples.

\end{document}